\documentclass[aps,prx,superscriptaddress,twocolumn]{revtex4-1}
\usepackage{amsmath}
\usepackage{amsfonts}
\usepackage{graphicx}
\usepackage{hyperref}
\usepackage{epstopdf}
\usepackage{bm}
\newcommand{\mm}{\mathbf m}
\newcommand{\RR}{\mathbf R}
\newcommand{\euler}{C}
\newcommand{\half}{\frac{1}{2}}

\begin{document}
\title{Inertia and chiral edge modes of a skyrmion magnetic bubble}
\author{Imam Makhfudz}
\affiliation{Department of Physics and Astronomy, Johns Hopkins University, Baltimore, Maryland 21218, USA}
\author{Benjamin Kr{\"u}ger}
\affiliation{I. Institut f{\"u}r Theoretische Physik, Universit{\"a}t Hamburg, Jungiusstr. 9, 20355 Hamburg, Germany}
\author{Oleg Tchernyshyov}
\affiliation{Department of Physics and Astronomy, Johns Hopkins University, Baltimore, Maryland 21218, USA}

\begin{abstract}
The dynamics of a vortex in a thin-film ferromagnet resembles the motion of a charged massless particle in a uniform magnetic field. Similar dynamics is expected for other magnetic textures with a nonzero skyrmion number. However, recent numerical simulations revealed that skyrmion magnetic bubbles show significant deviations from this model. We show that a skyrmion bubble possesses inertia and derive its mass from the standard theory of a thin-film ferromagnet. Besides center-of-mass motion, other low energy modes are waves on the edge of the bubble traveling with different speeds in opposite directions.
\end{abstract}

\maketitle

Dynamics of topological defects is a topic of long-standing interest in magnetism. The attention to it stems from rich basic physics as well as from its connection to technological applications \cite{Chien07}. Theory of magnetization dynamics in ferromagnets well below the critical temperature is based on the Landau-Lifshitz equation \cite{Landau35} for the unit vector of magnetization $\mm(\mathbf r) = \mathbf M(\mathbf r)/M$,
\begin{equation}
\dot{\mm} = \gamma \mathbf B \times \mm + \alpha \mm \times \dot{\mm},
\label{eq:LLG}
\end{equation}
where $\gamma$ is the gyromagnetic ratio, $\alpha \ll 1$ is a phenomenological damping constant \cite{Gilbert04}, and the effective magnetic field is a functional derivative of the free energy,  $\mathbf B(\mathbf r) = - \delta U/\delta \mathbf M(\mathbf r)$. The latter includes local (e.g., exchange and anisotropy) as well as long-range (dipolar) interactions, thus making Eq.~(\ref{eq:LLG}) a nonlinear and nonlocal partial differential equation with multiple length and time scales solvable in only a few simple cases.  For example, translational motion of a rigid texture, $\mm = \mm(\mathbf r - \mathbf R(t))$, is fully parametrized by the texture's ``center of mass'' $\mathbf R$. For steady motion, $\mathbf R(t) = \mathbf V t$, the velocity can be obtained from Thiele's equation \cite{Thiele73} expressing the balance of gyrotropic, conservative, and viscous forces:
\begin{equation}
\mathbf G \times \dot{\mathbf R} + \mathbf F - D \dot{\mathbf R} = 0.
\label{eq:Thiele}
\end{equation}
Here $\mathbf G$ is a gyrocoupling vector, $\mathbf F = -\partial U/\partial \mathbf R$ is the net conservative force, and $D$ is a dissipation tensor. A rigid texture moves like a massless particle with electric charge in a magnetic field and an external potential through a viscous medium. If $\mathbf G \neq 0$, the ``Lorentz force'' greatly exceeds the viscous drag. We thus ignore dissipation.

Although Eq.~(\ref{eq:Thiele}) was derived for steady motion, Thiele anticipated that it could serve as a good first approximation in more general situations. Indeed, his equation describes very well the dynamics of vortices in thin ferromagnetic films \cite{guslienko:8037, Science.304.420, PhysRevLett.95.167201, guslienko:067205, PhysRevB.76.224426}. In this case, the gyrocoupling vector $\mathbf G = (0,0,\mathcal G)$ is proportional to a topological invariant known as the skyrmion charge $q = (1/4\pi) \int dx \, dy \, \mm \cdot (\partial_x \mm \times \partial_y \mm)$, the film thickness $t$, and the density of angular momentum $M/\gamma$; to wit, $\mathcal G = 4\pi q t M/\gamma$. A vortex has $q = \pm 1/2$ and thus $\mathbf G \neq 0$. In a parabolic potential well, $U(X,Y) = \mathcal K(X^2+Y^2)/2$, it moves in a circle at a frequency $\omega = \mathcal K/\mathcal G$. 

\begin{figure}
\includegraphics[width=0.95\columnwidth]{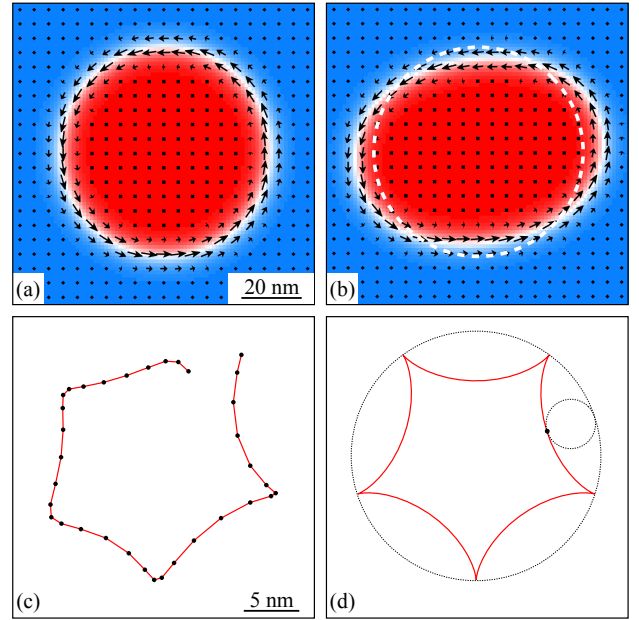}
\caption{A circular magnetic bubble with skyrmion number $q=-1$. Out-of-plane magnetization is blue for $M_z>0$ and red for $M_z<0$; in-plane magnetization on the domain wall is shown by arrows.  (a) Equilibrium. (b) Elliptic deformation. The dashed line marks the equilibrium position of the domain wall. (c) The trajectory of the bubble's center observed by Moutafis \emph{et al}. Dots mark positions evenly spaced in time. (d) A 5-cusped hypocycloid with its directing and generating circles. }
\label{fig:bubble}
\end{figure}

Similar behavior is expected for other topologically nontrivial textures, e.g., magnetic bubbles in thin films with magnetization normal to the plane of the film \cite{Malozemoff:1979, Science.272.1782, Science.284.1969}. A bubble is a circular domain with $m_z<0$ surrounded by a domain with $m_z>0$, or vice versa, Fig.~\ref{fig:bubble}(a) \cite{PhysRevB.71.060405}. Recently discovered skyrmion crystals \cite{Nature.442.797, Science.323.915}, particularly those which are found in thin films \cite{Nature.465.901, PhysRevLett.108.267201}, are periodic arrays of magnetic bubbles with the same skyrmion charge $q = \pm 1$ \cite{PNAS.109.8856}. Zang \emph{et al.} \cite{PhysRevLett.107.136804} modeled skyrmions in these structures as massless particles with Thiele's dynamics (\ref{eq:Thiele}). However, numerical simulations of Moutafis \emph{et al.} \cite{Moutafis:224429} revealed that this model fails badly for an isolated skyrmion bubble. In a parabolic potential, the trajectory of the bubble's center was not a circle, but ``roughly a pentagon,'' Fig.~\ref{fig:bubble}(c). 

The main goal of this Letter is to derive the correct dynamical model of a skyrmion magnetic bubble. We first introduce our model phenomenologically and then derive it from the standard theory of a thin-film ferromagnet. This allows us to characterize not only the center-of-mass motion of the bubble but also the dynamics of its shape within the same framework. 

\emph{Phenomenology.} The puzzling trajectory of the bubble's center is readily reproduced if we endow a skyrmion bubble with inertial mass $\mathcal M$:
\begin{equation}
- \mathcal M\ddot{\RR} + \mathcal G \times \dot{\RR} - \mathcal K \RR = 0, 
\label{eq:Thiele-M}
\end{equation}
The new equation of motion is of the second order in $d/dt$ and has two circular modes with frequencies 
\begin{equation}
\omega_{\pm} = -\mathcal G/2 \mathcal M \pm \sqrt{(\mathcal G/2\mathcal M)^2 + \mathcal K/\mathcal M}. 
\label{eq:bubble-COM}
\end{equation}
A particle with zero initial velocity follows a hypocycloid. If $\omega_-/\omega_+ = -4$, the hypocycloid has 5 cusps and indeed resembles a pentagon, Fig.~\ref{fig:bubble}(d). We repeated the simulations of Moutafis \emph{et al.} and found that the motion of the bubble is described with good accuracy by a superposition of two underdamped modes with eigenfrequencies $\omega/2\pi = 0.97$ GHz and $-4.27$ GHz \cite{supmat}. 

To understand the origin of inertia, we shift attention from the center of the bubble, where nothing is happening, to its boundary, a domain wall defined as a line $y(x)$ where $M_z(x,y) = 0$. A nearly circular domain wall is conveniently parametrized in polar coordinates $(r, \phi)$:  
\begin{equation}
r(\phi) = \bar{r} + \sum_{m} r_m e^{i m \phi}.
\label{eq:r-phi}
\end{equation}
The Fourier amplitudes $r_m = r^*_{-m}$ describe waves with wavenumbers $k = m/\bar{r}$ traveling along the circular edge: $r_0$ is the breathing mode, $r_1 = (X-iY)/2$ encodes the location of the center of mass, $r_2$ parametrizes elliptical deformations, Fig.~\ref{fig:bubble}(b), and so on.  On the domain wall, magnetization lies in the plane of the film, $\mm = (\cos{\psi}, \sin{\psi},0)$. For a circular wall in equilibrium, $\mm$ points along the direction of the wall, $\psi = \phi \pm \pi/2$, Fig.~\ref{fig:bubble}(a). More generally,
\begin{equation}
\psi(\phi) = \phi \pm \pi/2 + \sum_{m} \psi_m e^{i m \phi}.
\label{eq:psi-phi}
\end{equation}
The fields $r(\phi)$ and $\psi(\phi)$ are coupled to each other, and so are their harmonics $r_m$ and $\psi_m$. Integrating out $\psi_1$ generates kinetic energy for the center of mass. 

\emph{Theory.} We derive the dynamics of transverse fluctuations of a Bloch domain wall, first for a straight wall and then for a circular one. To this end, we employ a method of collective coordinates generalizing Thiele's approach beyond steady motion \cite{tretiakov:127204}. The Lagrangian formalism allows us to easily integrate out the hidden degree of freedom---in-plane magnetization---in favor of the more evident transverse motion. 

An evolving magnetic texture $\mm(\mathbf r,t)$ can be parametrized by a (potentially infinite) set of collective coordinates $\bm\xi(t) = \{\xi_1(t), \xi_2(t), \ldots\}$. Their equations of motion are similar to Thiele's equation (\ref{eq:Thiele}):
\begin{equation}
G_{ij} \dot{\xi}_j + F_i - D_{ij} \dot{\xi}_j = 0, 
\label{eq:Thiele-xi}
\end{equation}
with generalized forces $F_i = -\partial U/\partial \xi_i$, gyrotropic coefficients $G_{ij} = -G_{ji}$, and viscosity coefficients $D_{ij} = D_{ji}$.  Eq.~(\ref{eq:Thiele-xi}) can be obtained from a Lagrangian $L = \mathbf A \cdot \dot{\bm \xi} - U(\bm\xi)$, where $\mathbf A(\bm \xi)$ is a gauge potential with curvature $\partial A_j/\partial \xi_i - \partial A_i/\partial \xi_j = G_{ij}$ \cite{clarke:134412}. The gauge term contributes to the action $S = \int L \, dt$ a time-independent piece $\int \mathbf A \cdot d \bm \xi$ known as Berry's geometric phase. 

\emph{Straight domain wall.} We first consider the dynamics of a domain wall stretched along the $x$-axis, $y(x,t) \approx 0$, $\psi(x,t) \approx 0$, from $x=0$ to $x=\ell$. Its Lagrangian,
\begin{equation}
L[y,\psi] = \int_0^\ell dx \, g \dot{y}\psi - U[y,\psi],
\end{equation}
contains a gauge term with gyrotropic coupling $g = 2t M/\gamma$ \cite{ZNaturforschA.3.373, PhysRep.478.213}. The resulting equations of motion are $-g\dot{\psi} - \delta U/\delta y = 0$, $g\dot{y} - \delta U/\delta \psi = 0$ in the absence of dissipation. The in-plane magnetization is aligned with the wall in equilibrium, $\psi = y' \equiv \partial y/\partial x$; the cost of small misalignments is quadratic in $\psi - y'$, so 
\begin{equation}
L[y,\psi] = \int_0^\ell dx \, 
	\left[ 
		g \dot{y}\psi 
		- \kappa(\psi-y')^2/2
	\right]
- U[y],
\end{equation}
with the stiffness $\kappa$ to be determined. The field $\psi$ can be integrated out with the aid of its equation of motion, $g\dot{y} - \kappa(\psi-y') = 0$, to obtain a Lagrangian for transverse displacements,
\begin{equation}
L[y] = \int_0^\ell dx \,
	\left(
		\rho \dot{y}^2/2 + g \dot{y} y'
	\right)
	- U[y],
\end{equation}
where $\rho = g^2/\kappa$ is the D{\"o}ring mass density \cite{ZNaturforschA.3.373, PhysRep.478.213}. 
 
Potential energy of a domain wall can be split into local and long-range contributions. The local term is proportional to the length of the wall and tension $\sigma$:  
\begin{equation}
U_l[y] 
	= \int_0^\ell \sigma \sqrt{dx^2 + dy^2} 
	\approx U_l[0] + \int_0^\ell dx \, \sigma y'^2/2, 
\label{eq:U-local}
\end{equation}
where $U_l[0] = \sigma \ell$ is the energy of a straight wall. Thus
\begin{equation}
L[y] = \int dx 
	\left(
		\rho \dot{y}^2/2 + g \dot{y} y' - \sigma y'^2/2 
	\right) 
	- U_{nl}[y].
\end{equation}
Neglecting for the moment the nonlocal term $U_{nl}[y]$, we obtain a wave equation, $\rho \ddot{y} + 2g \dot{y}' - \sigma y'' = 0$, with waves traveling left and right at \emph{different speeds}---cf. Eq.~(\ref{eq:bubble-COM}):
\begin{equation}
\omega = \rho^{-1}\left(-gk \pm \sqrt{g^2 k^2 + \sigma\rho k^2}\right). 
\label{eq:dispersion-local}
\end{equation}
In the limit $\sigma \ll g^2/\rho = \kappa$, the slow wave has velocity $v_1 \sim \sigma/2g$ that is insensitive to inertia; in-plane magnetization adiabatically aligns with the direction of the wall. The fast mode with velocity $v_2 \sim - 2g/\rho$ involves oscillations of in-plane magnetization out of phase with those of the direction of the wall. 

A minimal model of a thin-film ferromagnet with out-of-plane magnetization includes exchange coupling $A$ and easy-axis anisotropy $K$ strong enough to overcome the dipolar shape anisotropy: the ``quality factor'' $Q = 2K/\mu_0 M^2$ must exceed 1. In this model \cite{supmat}, a domain wall has the width $\lambda/\sqrt{Q-1}$, where $\lambda = \sqrt{2A/\mu_0M^2}$ is the exchange length, and tension $\sigma = 2 \mu_0 M^2 t \lambda \sqrt{Q - 1}$. The coupling between the in-plane magnetization and the direction of the wall is $\kappa = \sigma/(Q-1)$. By using the material parameters characteristic of FePt with thickness $t = 32$ nm \cite{Moutafis:224429}, we obtain $v_1 = 370$ m/s and $v_2 = -2100$ m/s in the local model. 

\begin{figure}
\includegraphics[width=0.95\columnwidth]{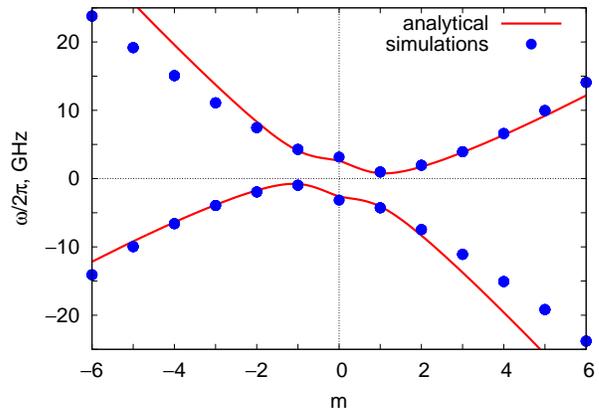}
\caption{The spectrum of spin waves $\omega/2\pi$ on a circular domain wall: numerical simulation (circles), model calculations for a straight wall (\ref{eq:dispersion-nonlocal}) with $k = m/\bar{r}$ (line).}
\label{fig:spectra}
\end{figure}

The nonlocal part of the potential energy comes from long-range dipolar interactions. A domain wall in a thin ferromagnetic film produces a stray magnetic field whose energy can be written as a double line integral along the domain wall \cite{PhysRevA.46.4894, Kashuba93},
\begin{equation}
U_{nl}[\mathbf r] = -\frac{\sigma_d}{2} \int \frac{d \mathbf r_1 \cdot d \mathbf r_2}{|\mathbf r_1 - \mathbf r_2|},
\label{eq:U-nonlocal}
\end{equation}
where $\sigma_d = \mu_0 M^2 t^2/\pi$ is ``dipolar tension.'' This expression diverges at both short and long length scales and thus requires both short and long-distance cutoffs (provided by the film thickness and the wall length). The local potential energy (\ref{eq:U-local}) may be absorbed into the nonlocal part (\ref{eq:U-nonlocal}) at the expense of renormalizing the short-distance cutoff. Expanding Eq.~(\ref{eq:U-nonlocal}) to the second order in $y$ yields the following result: 
\begin{equation}
U_{nl}[y] = U_{nl}[0] +  \ell \int_0^\infty \frac{dk}{2\pi} \, \sigma_d  k^2 \ln{(ka)} \, |y_k|^2,
\label{eq:U-nl-Fourier}
\end{equation}
where $y_k = \ell^{-1/2} \int_0^\ell dx \, y(x) e^{-ikx}$, $U_{nl}[0] \sim - \sigma_d \ell \ln{(\ell/a)}$ is the energy of a straight domain wall, $a = (t/2)e^{\euler - 1 + \sigma/\sigma_d}$ is a short-distance length scale, and $\euler = 0.577 \ldots$ is the Euler constant. The wave stiffness $\sigma_d k^2 \ln{(ka)}$ is negative for $k < 1/a$, which means that a straight domain wall is unstable against small deformations. This is the fingering instability occurring in systems with long-range interactions \cite{PhysRevA.46.4894, Science.267.467}. It can be prevented by placing the domain wall in a strip of finite width $w$. Repulsion from the edges, mediated by a stray magnetic field, increases the wave stiffness by a $k$-independent term $4\sigma_d/w^2$. The wave frequencies are
\begin{equation}
\omega = \rho^{-1}
	\left(
		-gk \pm \sqrt{g^2k^2 + \sigma_d\rho \left( k^2\ln{|ka|} + 4/w^2 \right)}
	\right).
\label{eq:dispersion-nonlocal}
\end{equation}
The frequency spectrum (Fig.~\ref{fig:spectra}) has two significant changes from the local model (\ref{eq:dispersion-local}): a gap opens up; the band bottom is shifted to $k_0 = e^{-1/2} a^{-1}$. 

\emph{Magnetic bubble.} The dynamics of a circular domain wall is derived along similar lines. For a wall of given shape (\ref{eq:r-phi}), the in-plane magnetization tends to align itself with the wall, $\psi_\mathrm{eq}(\phi) = \phi \pm\pi/2  - \bar{r}^{-1}\partial r/\partial\phi$. The $\psi$-dependent terms in the Lagrangian density are thus $g \dot{r} \psi - \kappa (\psi - \psi_\mathrm{eq})^2/2$. The Lagrangian of the Fourier modes (\ref{eq:r-phi}) and (\ref{eq:psi-phi}) is
\begin{equation}
L[r,\psi] = 2\pi \bar{r} \sum_m 
	\left(
		g \dot{r}_m \psi_m^* 
		- \frac{\kappa|\psi_m + i m r_m/\bar{r}|^2}{2}
	\right)
	- U[r].
\end{equation}
Integrating out $\psi$ yields a Lagrangian for $r$ containing both kinetic energy and a Berry-phase term: 
\begin{equation}
L[r] = \sum_{m} 
	\left(
		\pi \bar{r} \rho |\dot{r}_m|^2
		- 4\pi m i \, g \, r_m^* \dot{r}_m  
	\right) 
	- U[r].
\end{equation}

We thus arrive at a Lagrangian for the center-of-mass mode $r_1 = (X-iY)/2$: 
\begin{equation}
L(X,Y) = \mathcal M (\dot{X}^2 + \dot{Y}^2)/2 + \mathcal G \dot{X} Y - \mathcal K (X^2+Y^2)/2.
\label{eq:L-COM}
\end{equation}
It yields the anticipated inertial dynamics of a magnetic bubble (\ref{eq:Thiele-M}). The gyrotropic constant, $\mathcal G = 2\pi g = 4\pi tM/\gamma = 2.29 \times 10^{-12}$ J s/m$^2$, depends only on the topology of the bubble (here the skyrmion number $q=1$) and on the area density of angular momentum $t M/\gamma$; therefore, it can be taken at face value. The spring constant $\mathcal K$ comes from magnetostatic repulsion between the domain wall and the edge of the disk; its calculated value, $\mathcal K_\mathrm{calc} = 0.020$ J/m$^2$ for $\bar{r} = 37$ nm, should also be reliable. The weakest link in our theory is the mass term
$\mathcal M_\mathrm{calc} = \pi \bar{r} \rho = \pi \bar{r} g^2/\kappa = 5.4 \times 10^{-23}$ kg. It depends on the coupling $\kappa$ between in-plane magnetization and the direction of the wall and thus requires an accurate model of the domain wall. Alternatively, the basic constant $\mathcal G$ can be combined with the measured frequencies, $\omega_\mathrm{sim}/2\pi = 0.97$ GHz and -4.27 GHz, to obtain $\mathcal K_\mathrm{sim} = 0.018$ J/m$^2$ and $\mathcal M_\mathrm{sim} = 1.11 \times 10^{-22}$ kg. As expected, we find a good match between the calculated and simulated values of the spring constant $\mathcal K$, but the $\mathcal M$ values differ by a factor of 2. 

We have performed numerical simulations to measure the frequency spectrum of waves with higher azimuthal numbers $m$. We used the same geometry and material parameters as Moutafis \emph{et al.}: a FePt disk of radius $R = 80$ nm, thickness $t = 32$ nm, magnetization $M = 10^6$ A/m, exchange constant $A = 10^{-11}$ J/m, easy-axis anisotropy $K = 1.3 \times 10^6$ J/m$^3$, and gyromagnetic ratio $\gamma = 1.75 \times 10^{11}$ s A/kg. These give the quality factor $Q = 2.06$ and exchange length $\lambda = 4.0$ nm. We utilized micromagnetic simulator OOMMF \cite{oommf} in the two-dimensional regime with a unit cell of 1.25 nm. The equilibrium radius of the bubble was $\bar{r} = 37$ nm. The free motion of the $m$th harmonic of $\rho(\phi)$ was fitted by a sum of two underdamped components. The extracted frequencies are shown in Fig.~\ref{fig:spectra}. The theoretical curve is the straight-wall spectrum (\ref{eq:dispersion-nonlocal}) with $k = m/\bar{r}$. We used the mass density extracted from the simulation of the $m=1$ mode, $\rho = \mathcal M_\mathrm{sim}/\pi \bar{r}$, and set the effective width $w$ equal to the disk radius. The theory works quite well for the slow mode, less so for the fast one.

\emph{Discussion.} The widely used Thiele's equation (\ref{eq:Thiele}) predicts that magnetic textures with a skyrmion number $q \neq 0$ behave as massless particles moving in a uniform magnetic field and an external potential. Although this approach works very well for magnetic vortices ($q=\pm 1/2$), it fails for skyrmion magnetic bubbles ($q = \pm 1$). Here we have shown that a skyrmion bubble behaves as a massive object and have explained the origin of its mass. A skyrmion bubble possesses additional modes, which are best viewed as transverse fluctuations of its edge. These waves are chiral, i.e., they propagate with different speeds in opposite directions. The non-reciprocal wave propagation is expected in other geometries, e.g., striped and labyrinthine domains.

In a skyrmion crystal, the discrete modes of a bubble turn into excitation branches. The slow and fast $m=1$ modes give rise to the magnetophonon and cyclotron branches \cite{PhysRevB.84.214433}; the cyclotron frequency $\omega = \mathcal{G/M}$ is in the GHz range. The $m = 0$ breathing mode has been seen in numerical simulations \cite{PhysRevLett.108.017601}. The breating mode and one of the $m=1$ modes have been found in ${\mathrm{Cu}}_{2}{\mathrm{OSeO}}_{3}$ \cite{PhysRevLett.109.037603}. Branches with higher $m$ may also be detectable.

\emph{Acknowledgments.} We thank Stavros Komineas for helpful comments on the manuscript. This work was supported in part by the US National Science Foundation under Award No. DMR-1104753.

\bibliographystyle{apsrev4-1}
\bibliography{micromagnetics,skyrmions,supmat}

%%%%%%%%%%%%%%%%%%%%%%%%%%%%%%%%%%%%%%%%%%%%%%%%%%%%%%%%%%%%%%%%%%%%%%%%%%%%%%%%%%%%%%%%%%%%%%%%%%%%%%%%%%%%
% Supplemental Material
%%%%%%%%%%%%%%%%%%%%%%%%%%%%%%%%%%%%%%%%%%%%%%%%%%%%%%%%%%%%%%%%%%%%%%%%%%%%%%%%%%%%%%%%%%%%%%%%%%%%%%%%%%%%

\newpage
\setcounter{figure}{0}
\appendix

\begin{center}
{\large \bf Supplemental Material}
\end{center}

\section{Center-of-mass motion}

\begin{figure}[b]
\includegraphics[width=0.95\columnwidth]{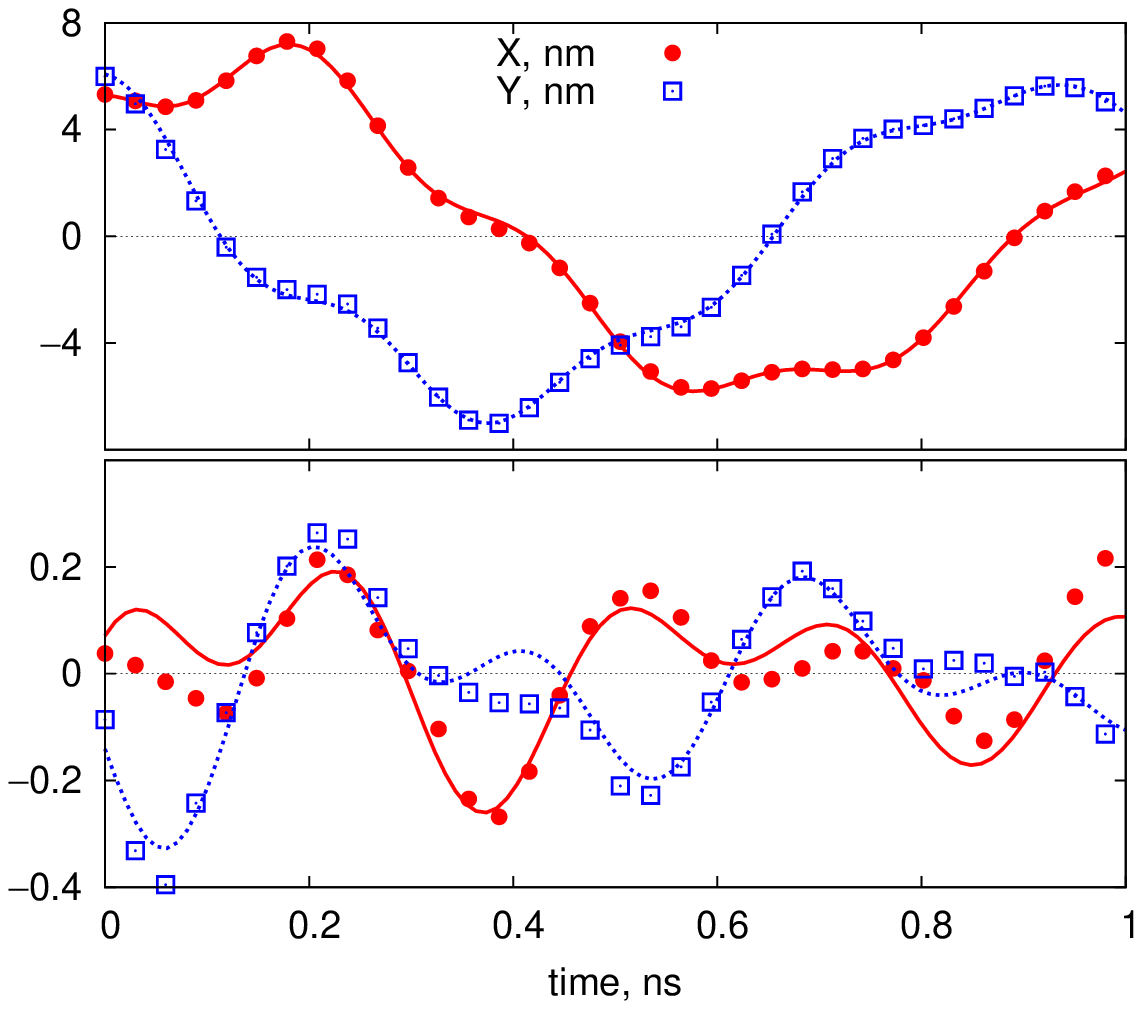}
\caption{Top panel: The displacement of the bubble's center of mass $(X,Y)$ as a function of time (points) and the best fit to the two-mode approximation (\ref{eq:r1-t}) (lines). Bottom panel: The difference between the data and the best fit (points) and the best fit to a superposition of two spiral motions with frequencies $\omega_{+1} \pm \omega_0$ (lines). The vertical scales in the two panels differ by a factor of 30.}
\label{fig:COM}
\end{figure}

The motion of the center of mass $(X,Y)$ of a magnetic bubble is conveniently represented by a complex variable $r_1 = (X-iY)/2$. The general motion of a bubble with inertial mass $\mathcal M$ and a gyrotropic coefficient $\mathcal G$ in a parabolic potential with stiffness $\mathcal K$ is a superposition of two spiral motions: 
\begin{equation}
r_1(t) = \sum_{m = \pm 1} a_{m} e^{i\omega_{m} t - \Gamma_{m} t}. 
\label{eq:r1-t}
\end{equation}
For weak dissipation, $\Gamma_{m} \ll \omega_{m}$, 
\begin{equation}
\omega_{\pm 1} \approx -\frac{\mathcal G}{2 \mathcal M} 
	\pm \sqrt{\left( \frac{\mathcal G}{2 \mathcal M} \right)^2 + \frac{\mathcal K}{\mathcal M}},
\quad
\frac{\Gamma_{+1}}{\Gamma_{-1}} \approx \left| \frac{\omega_{+1}}{\omega_{-1}} \right|.
\end{equation}
The top panel of Fig.~\ref{fig:COM} shows the best fit to Eq.~(\ref{eq:r1-t}) with $\omega_{+1}/2\pi = 0.97$ GHz, $\Gamma_{+1} = 0.25$ ns$^{-1}$, $\omega_{-1}/2\pi = -4.27$ GHz, and $\Gamma_{-1} = 1.0$ ns$^{-1}$. 

A small systematic deviation between the best-fit line and the data is plotted in the bottom panel of Fig.~\ref{fig:COM}. Much of it can be accounted for by a superposition of two spiral motions with frequencies $\omega_{+1} \pm \omega_0$, where $\omega_0/2\pi = 3.15$ GHz is the eigenfrequency of the bubble's breathing mode. This is likely the effect of an anharmonic coupling between the two modes. 

We thus conclude that any unidentified modes contribute no more than 0.1 nm in amplitude, or about 1 per cent, to the center-of-mass motion. 

\section{Locating the domain wall}

We define the domain wall as a line of points in the plane of the film where out-of-plane magnetization $M_z(x,y)$ vanishes. Because our numerical simulations were done on a discrete lattice, we needed an algorithm to extract a continuous line from discrete data points. The location of the domain wall was determined in two steps: we first identified a discrete set of points with $M_z=0$; we then fit their positions to a line, Eq.~(5) in the main text.

\begin{figure}
\includegraphics[width = 0.6\columnwidth]{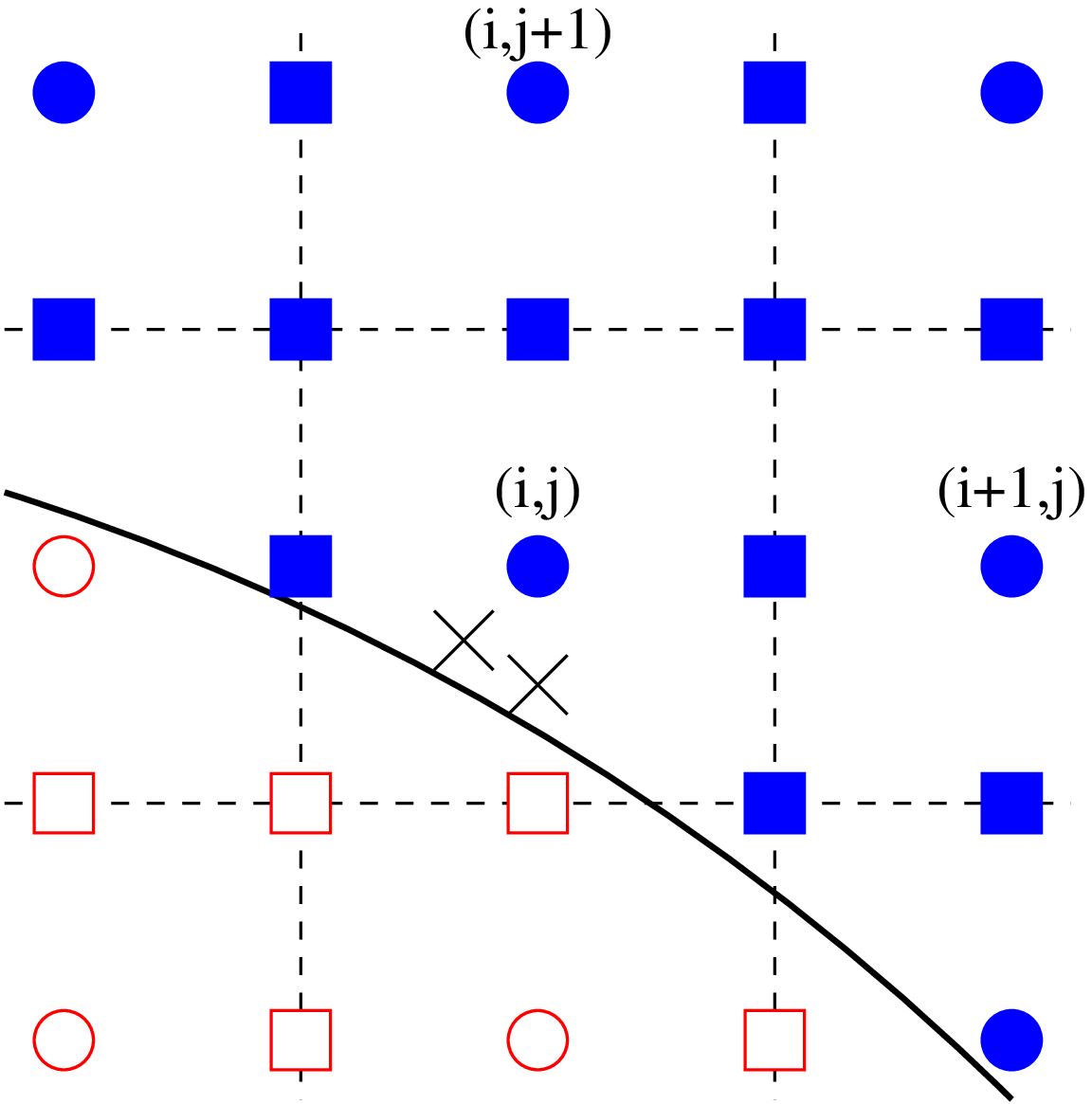}
\caption{Tracing a domain wall (solid line).  Circles denote the centers of lattice cells in the simulation. Squares are points at cell boundaries (dashed lines).  Crosses are locations of the domain wall obtained by linear extrapolation. Open red symbols mean $M_z>0$, filled blue symbols mean $M_z<0$.}
\label{find_boundary}
\end{figure}

The first step is illustrated in Fig.~\ref{find_boundary}. The magnetization at the boundary of a cell is determined by interpolation with the magnetization at the center of the current cell and its nearest neighbors as follows:
\begin{eqnarray*}
\mathbf M_{i,j+\half} & = & \left( \mathbf M_{i,j} + \mathbf M_{i,j+1} \right)/2, \\
\mathbf M_{i+\half,j} & = & \left( \mathbf M_{i,j} + \mathbf M_{i+1,j} \right)/2, \\
\mathbf M_{i+\half,j+\half} & = & \left( \mathbf M_{i,j} + \mathbf M_{i,j+1} + \mathbf M_{i+1,j} + \mathbf M_{i+1,j+1} \right)/4.
\end{eqnarray*}
If the wall intersects the cell then the sign of $M_z$ at one or more points at its boundary differs from that at the center of the cell. At the crudest level, we could use the centers of intersected cells as a proxy for the location of the domain wall. To refine this result, we used a linear interpolation for $M_z$ between the center of a cell and the points on its boundary to find the locations with $M_z = 0$. The refinement reduced the discretization noise by a factor of 70, which was particularly important for modes with higher azimuthal numbers, which had small amplitudes. 

\section{Local interactions} 

Tension $\sigma$ and alignment stiffness $\kappa$ come from the local portion of the energy functional. Consider a  ferromagnetic film of thickness $t$ with two domains of magnetization, $\mm = (0,0,1)$ and $\mm = (0,0,-1)$, separated by a straight domain wall of length $\ell$ along the $x$-axis.  The magnetization field can be parametrized as 
\[
\mm(y) = (\sin{\theta}\cos{\psi}, \sin{\theta}\sin{\psi}, \cos{\theta})
\]
with $\theta = \theta(y)$ and $\psi = \operatorname{const}$. In the state of lowest energy, in-plane magnetization points along the wall, $\psi = 0$ (Bloch wall). The energy contains three terms: exchange 
\[
U_\mathrm{exchange} = 
A t \int d^2r \, (\partial_i \mathbf m) \cdot (\partial_i \mathbf m) 
	= A t\ell \int_{-\infty}^\infty dy \left( \frac{d\theta}{dy} \right)^2, 
\]
easy-axis anisotropy  
\[
U_\mathrm{anisotropy} = 
-K t \int d^2r \,  m_z^2 = -K t\ell  \int_{-\infty}^\infty dy\, \cos^2{\theta},
\] 
and the energy of the magnetic field, which can be evaluated as the Coulomb interaction of magnetic charges at the top and bottom surfaces of the film with area densities $\mp M_z(\mathbf r)$: 
\[
U_\mathrm{dipolar} =
	\frac{\mu_0}{4\pi} \int d^2 r_1 \, d^2 r_2 \, 
V(\mathbf r_1 - \mathbf r_2) M_z(\mathbf r_1) M_z(\mathbf r_2) 
\]
with the interaction kernel 
\[
V(\mathbf r) = \frac{1}{r} - \frac{1}{\sqrt{r^2+t^2}} \equiv \mathbb{V}(r).
\]
For future convenience, we extend the definition of $\mathbb V(x)$ to negative values of the argument so that $\mathbb V(-x) = \mathbb V(x)$: 
\begin{equation}
\mathbb{V}(x) = \frac{1}{|x|} - \frac{1}{\sqrt{x^2+t^2}} 
= \lim_{\epsilon \to 0} \frac{1}{\sqrt{x^2+\epsilon^2}} - \frac{1}{\sqrt{x^2+t^2}}.
\label{eq:Vbb}
\end{equation}
The interaction kernel $V(\mathbf r)$ has a peak with a characteristic width of the order of the film thickness $t$. If magnetization varies slowly on that length scale, we may approximate 
$V(\mathbf r) \approx 2 \pi t \delta(\mathbf r)$. Then the dipolar energy assumes a local form:
\begin{equation}
U_\mathrm{dipolar} \approx t \int d^2r \, \frac{\mu_0 M_z^2}{2}
	= \frac{\mu_0M^2 t\ell}{2} \int_{-\infty}^\infty dy \, \cos^2{\theta}.
\label{eq:dipolar-local}
\end{equation}
Put another way, we make a local approximation for the magnetic field inside the film, $H_z(\mathbf r) \approx -M_z(\mathbf r)$, whose energy density $\mu_0 H^2/2 \approx \mu_0 M_z^2/2$. 

The energy cost of a domain wall is thus
\begin{equation}
U = \ell t \int_{-\infty}^{\infty} dy 
	\left[
		A {\theta'}^2 + (K - \mu_0 M^2/2) \sin^2{\theta} 
	\right].
\label{eq:energy-local}
\end{equation}
Minimization of this energy yields the domain-wall profile 
\begin{equation}
\cos{\theta(y)} = \tanh{\left( 
		\frac{y}{\lambda} \sqrt{Q-1} 
	\right)},
\end{equation}
where $\lambda = \sqrt{2A/\mu_0 M^2}$ is the exchange length and $Q = 2K/\mu_0 M^2 > 1$ is the ``quality factor.'' The minimized energy per unit length gives line tension 
\begin{equation}
\sigma = U/\ell = 2 \mu_0 M^2 t \lambda \sqrt{Q - 1}. 
\end{equation}

When in-plane magnetization deviates from the direction of the wall, it creates bulk magnetic charges with volume density $-\nabla \cdot \mathbf M = - d M_y/dy$. This induces an additional magnetic field $H_y \approx - M_y$ and thus generates additional energy density $\mu_0 H_y^2/2 \approx (\mu_0 M^2/2) \sin^2{\theta} \sin^2{\psi}$. The domain-wall energy per unit length increases to
\begin{equation}
U(\psi)/\ell = 2 \mu_0 M^2 t \lambda \sqrt{Q - 1 + \sin^2{\psi}} \sim \sigma + \kappa \psi^2/2
\end{equation}
for small angles $\psi$.  The strength of coupling between the azimuthal angle and the direction of the wall is
\begin{equation}
\kappa = \frac{1}{\ell}\left. \frac{d^2 U}{d \psi^2} \right|_{\psi=0} = \frac{\sigma}{Q-1}. 
\end{equation}

\section{Long-range interaction}

Equation (\ref{eq:dipolar-local}) captures the local part of dipolar energy, whose density is determined by the local value of magnetization. Inhomogeneities in magnetization produce a stray magnetic field that gives rise to a nonlocal component, 
\begin{equation}
U_\mathrm{stray} = \frac{\mu_0}{4\pi} \int d^2r_1 \, d^2r_2 \, 
	G(\mathbf r_1 - \mathbf r_2) \, 
	\nabla M_z(\mathbf r_1) \cdot \nabla M_z(\mathbf r_2),
\label{eq:U-gradients}
\end{equation}
where $\mathbf r = (x,y)$ and $\nabla = (\partial_x, \partial_y)$ are two-dimensional vectors. 
The kernel is defined as a solution to the equation $\nabla^2 G(\mathbf r) = -V(\mathbf r)$ and has the explicit form 
\begin{equation}
G(\mathbf r) =	\sqrt{r^2+t^2} - r - t \operatorname{arsinh}(t/r) \equiv \mathbb{G}(r).
\label{eq:G-exact}
\end{equation}
As we did for $\mathbb V(x)$, we extend the definition of $\mathbb{G}(x)$ to $x < 0$ so that $\mathbb{G}(-x) = \mathbb{G}(x)$. Note that 
\begin{equation}
\mathbb G''(x) + \mathbb G'(x)/x = - \mathbb V(x).
\label{eq:Gbb-Vbb}
\end{equation}

For an infinitely sharp domain wall, Eq.~(\ref{eq:U-gradients}) reduces to a double line integral,
\begin{equation}
U_\mathrm{stray}[\mathbf r] 
	= \frac{\mu_0 M^2}{\pi} \int d \mathbf r_1 \cdot d \mathbf r_2 \, G(\mathbf r_1 - \mathbf r_2).
\label{eq:U-nonlocal-exact}
\end{equation}
The asymptotic form of the kernel (\ref{eq:G-exact}) is $G(\mathbf r) \sim -t^2/2r$ for $r \gg t$. Hence 
\begin{equation}
U_\mathrm{stray}[\mathbf r] 
	\approx -\frac{\sigma_d}{2} \int \frac{d \mathbf r_1 \cdot d \mathbf r_2}{|\mathbf r_1 - \mathbf r_2|}
\label{eq:U-nonlocal-approx}
\end{equation}
with ``dipolar tension'' 
\begin{equation}
\sigma_d = \frac{\mu_0 M^2t^2}{\pi}. 
\label{eq:sigma-d}
\end{equation}

The simplified version of the stray-field interaction (\ref{eq:U-nonlocal-approx}) is logarithmically divergent at short distances. One way to handle the divergence is to impose a short-distance cutoff, $|\mathbf r_1 - \mathbf r_2| > b$, where $b$ is a length scale of the order of the film thickness $t$. The energy of a straight wall due to its stray field, computed with the exact expression (\ref{eq:U-nonlocal-exact}), is
\[
U_\mathrm{stray} \sim -\sigma_d \ell 
	\left[
		\ln{(2\ell/t)} + 1/2
	\right],
\]
whereas the simplified version (\ref{eq:U-nonlocal-approx}) with a cutoff yields 
\begin{equation}
U_\mathrm{stray} \sim -\sigma_d \ell 
	\left[
		\ln{(\ell/b)} - 1
	\right].
\label{eq:U-straight-cutoff}
\end{equation}
The two expressions agree if we choose the cutoff to be $b = (t/2) e^{-3/2}$.

Local contributions to the energy of the domain wall $\sigma \ell$ can be absorbed into the stray-field energy (\ref{eq:U-straight-cutoff}) at the expense of renormalizing the cutoff parameter, 
\begin{equation}
b = (t/2) e^{-3/2+\sigma/\sigma_d}.
\end{equation}

\section{Energy of transverse fluctuations}

We compute the energy of a nearly straight domain wall, $y(x) \approx 0$. The local part can be written as
\begin{equation}
U_\mathrm{loc}[y] = \int_0^\ell dx \frac{\sigma y'^2}{2} 
= \ell \int_0^\infty \frac{dk}{2\pi} \sigma k^2 |y_k|^2,
\label{eq:U-flucts-local}
\end{equation}
where $y_k = \ell^{-1/2} \int_0^\ell dx \, y(x) e^{-ikx}$ is the spatial Fourier transform of $y(x)$. 

The energy of the stray field (\ref{eq:U-nonlocal-exact}) can be expanded to $\mathcal O(y^2)$ with the aid of the following results: 
\[
d\mathbf r_1 \cdot d \mathbf r_2 = dx_1 \, dx_2 
	\left[
		1 + y'(x_1)  y'(x_2)
	\right],
\]
\[
G(\mathbf r) = \mathbb G(\sqrt{x^2+y^2}) 
	= \mathbb G(x) 
		+ \mathbb G'(x) \, y^2/2x
		+ \mathcal O(y^4).
\] 
The nonlocal portion of the energy of transverse fluctuations thus consists of two parts,
\begin{eqnarray*}
U_\mathrm{stray}[y] &=& \frac{\mu_0 M^2}{\pi} \int dx_1 \, dx_2 \, \mathbb G(x_1-x_2) y'(x_1) y'(x_2)
\\
	&+& \frac{\mu_0 M^2}{\pi} \int dx_1 \, dx_2 \, \frac{\mathbb G'(x_1-x_2)}{2(x_1-x_2)} (y_1-y_2)^2.
\end{eqnarray*}
After an integration by parts and a Fourier transform, we obtain 
\[
U_\mathrm{stray}[y] = \ell \int_0^\infty \frac{dk}{2\pi} A_k |y_k|^2,
\]
where 
\[
A_k = \frac{2 \mu_0 M^2 }{\pi} \int_{-\infty}^\infty dx \, [1-\cos{(kx)}] 
	\left[
		\mathbb G''(x) + \mathbb G'(x)/x
	\right].
\]
With the aid of Eqs.~(\ref{eq:Vbb}) and (\ref{eq:Gbb-Vbb}), we obtain
\[
A_k = -\frac{4\mu_0 M^2}{\pi} \left[
		K_0(kt) + \ln{(kt/2)} + \euler
	\right],
\]
where $K_0(x)$ is a modified Bessel function and $\euler = 0.577\ldots$ is the Euler constant. In the infrared limit, $kt \to 0$,
\[
A_k \sim \sigma_d k^2 
	\left[ 
		\ln{(kt/2)} + \euler - 1
	\right].
\]
Upon adding the local term (\ref{eq:U-flucts-local}), we obtain
\begin{equation}
U[y] \sim \ell \int_0^\infty \frac{dk}{2\pi} \sigma_d k^2 \ln{(ka)} |y_k|^2
\label{eq:U-nonlocal-asymptotic}
\end{equation}
with the short-distance scale 
\begin{equation}
a = (t/2)e^{\euler - 1 + \sigma/\sigma_d}. 
\label{eq:a}
\end{equation}

\end{document}